\documentclass[reprint, amsmath, amssymb, aps, prl]{revtex4-1}
\usepackage{graphicx}
\usepackage{amssymb}
\usepackage{amsmath}
\usepackage{setspace}
\usepackage{dcolumn}
\usepackage{newfloat}
\usepackage{epsfig}

\usepackage[colorlinks=true, citecolor=blue, linkcolor=blue, urlcolor=blue]{hyperref}

\usepackage{epstopdf}
\usepackage{bm}
\usepackage{float}
\usepackage{color}

\def\bea{\begin{eqnarray}}
\def\eea{\end{eqnarray}}
\def\be{\begin{equation}}
\def\ee{\end{equation}}

\usepackage{datetime}
\advance\currenthour by 0

\begin{document}

\title{Nonlinear Phonon Interferometry at the Heisenberg Limit}
\author{H. F. H. Cheung, Y. S. Patil, L. Chang, S. Chakram and M. Vengalattore}
\affiliation{Laboratory of Atomic and Solid State Physics, Cornell University, Ithaca, NY 14853}
\email{mukundv@cornell.edu}

\begin{abstract}
Interferometers operating at or close to quantum limits of precision have found wide application in tabletop searches for physics beyond the standard model, the study of fundamental forces and symmetries of nature and foundational tests of quantum mechanics. The limits imposed by quantum fluctuations and measurement backaction on conventional interferometers ($\delta \phi \sim 1/\sqrt{N}$) have spurred the development of schemes to circumvent these limits through quantum interference, multiparticle interactions and entanglement. A prominent example of such schemes, the so-called $SU(1,1)$ interferometer, has been shown to be particularly robust against particle loss and inefficient detection, and has been demonstrated with photons and ultracold atoms. Here, we realize a $SU(1,1)$ interferometer in a fundamentally new platform in which the interfering arms are distinct flexural modes of a millimeter-scale mechanical resonator. We realize up to 15.4(3) dB of noise squeezing and demonstrate the Heisenberg scaling of interferometric sensitivity ($\delta \phi \sim 1/N$), corresponding to a 6-fold improvement in measurement precision over a conventional interferometer. Our work extends the optomechanical toolbox for the quantum manipulation of macroscopic mechanical motion and presents new avenues for studies of optomechanical sensing and the nonequilibrium dynamics of multimode optomechanical systems. 
\end{abstract}

\maketitle

Interferometers are an indispensable metrological tool in the study of fundamental forces \cite{thorne1979, dimopoulos2007}, the search for physics beyond the standard model \cite{tino2014} and the measurement of fundamental constants \cite{cronin2009}. The realization that conventional interferometry is limited by quantum fluctuations and measurement backaction has led to the concept of the standard quantum limit (SQL) \cite{caves1980b}. This has spurred efforts to observe quantum effects in macroscopic interferometers \cite{caves1980} and to circumvent the SQL via entanglement \cite{leibfried2004, wasilewski2010}, multiparticle interactions \cite{napolitano2011} and quantum interference \cite{hammerer2009, tsang2012}. In a broader sense, these efforts have led to theoretical studies aimed at elucidating the metrological precision of a macroscopic quantum many-body system and the interplay between entanglement, many-body interactions, topology and nonlinearities \cite{boixo2008, negretti2008, hall2012, woolley2008}, as well as experimental efforts to investigate these questions in atomic, solid state and hybrid quantum systems.  

Optomechanical systems have emerged as a promising arena for the investigation of these foundational aspects of quantum metrology and the innovation of novel precision measurement technologies \cite{aspelmeyer2014}. The enormous range of size and mass, spanning the nanoscale to the macroscale, the long coherence times that compare favorably with those realized in atomic or solid state spin systems \cite{chakram2014, yuan2015, reinhardt2015, norte2015}, and the ability to cool, probe and control mechanical motion with radiation pressure have aided these efforts. While optomechanical interactions have thus far been mainly in the weak coupling regime, recent work has demonstrated the possibility of realizing strong, quantum-compatible nonlinear or multimode mechanical interactions via radiation pressure \cite{pontin2015}, geometric design \cite{mahboob2014} or reservoir engineering \cite{patil2015}. This raises prospects of manipulating macroscopic mechanical states in the quantum regime with techniques similar to those in quantum or atom optics.
\begin{figure*}[t]
\centering
\includegraphics[width=0.90\textwidth]{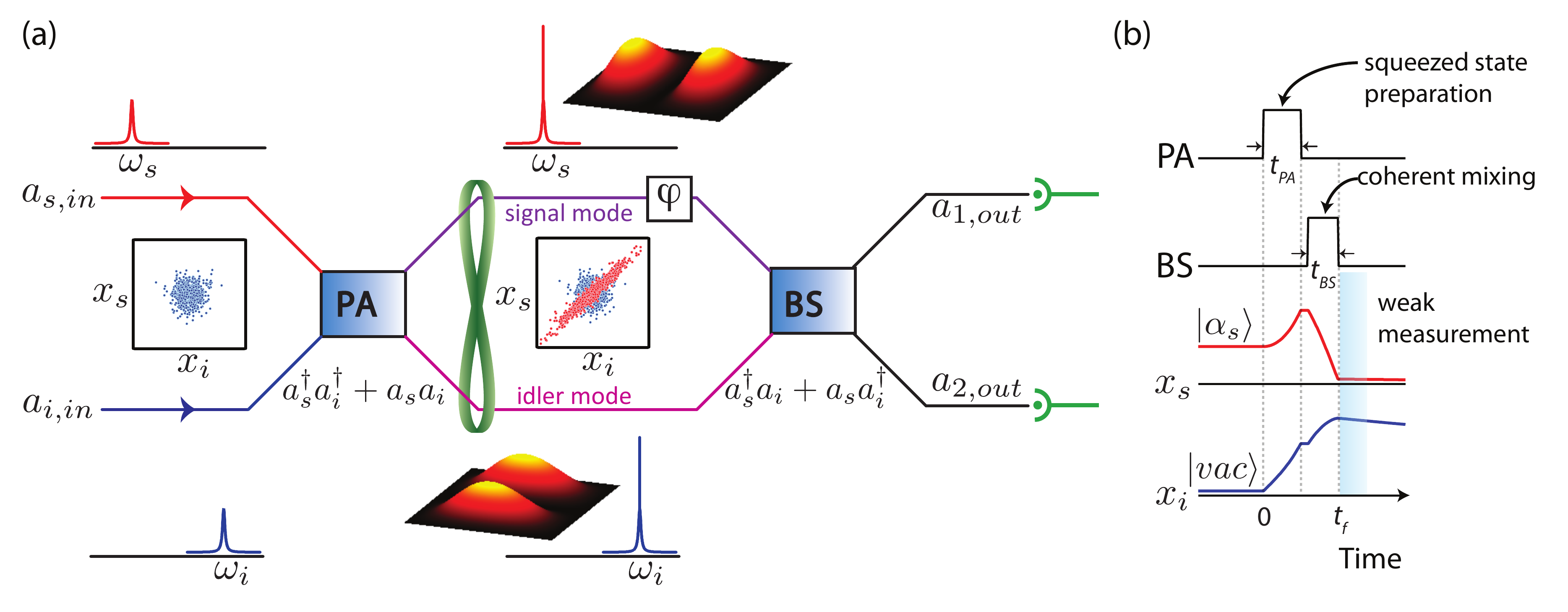}
\caption{\textbf{A $SU(1,1)$ phonon interferometer.} (a) The two arms of the interferometer are distinct mechanical modes at frequences $\omega_s$ and $\omega_i$. A parametric amplifier interaction (PA) between the two modes generates strong correlations between these modes. A phase shift of interest $\varphi$ is then imparted to the signal mode. A timed and pulsed beam splitter interaction (BS) between the modes coherently mixes the two correlated arms resulting in reduced quadrature noise at the outputs. (b) The timing sequence : The input to the interferometer is the coherent state $|\alpha_s,0\rangle$ prepared at $t<0$. The signal and idler get correlated during the parametric amplifier pulse for time $t_{PA}$. After a variable interaction period,  the two modes are coherently mixed by the beam splitter pulse for time $t_{BS}$, followed by a weak measurement of the output modes. }
\label{fig:figschem}
\end{figure*}

In this work, we realize a nonlinear phonon interferometer in a millimeter-scale mechanical resonator and 
demonstrate Heisenberg scaling of phase sensitivity with phonon number. By using quantum-compatible two-mode nonlinearities
to create strong correlations between the two modes of the interferometer, we demonstrate up to 15.4(3) dB of noise squeezing
and a 6-fold enhancement in measurement sensitivity over a conventional interferometer. The features of our nonlinear interferometer are accurately captured by a model applicable to a generic pair of parametrically coupled oscillators which also shows that the achievable noise reduction is, in principle, unbounded, enabling an unrestricted improvement in signal-to-noise ratio. 

A schematic of the nonlinear phonon interferometer is shown in Fig.~\ref{fig:figschem}. While this schematic highlights the nominal similarities to a Mach-Zehnder or a Ramsey interferometer, we note two key differences. First, the arms of the interferometer consist of two distinct mechanical modes of a silicon nitride (SiN) membrane resonator. The motion of these modes can be spectroscopically resolved and independently measured via an optical interferometer as described in previous work \cite{patil2015}. Unlike in the optical domain, the phonons in this interferometer are necessarily confined within a cavity, i.e. the mechanical resonator, and do not freely propagate. In this sense, the mechanical modes are more analogous to intracavity optical fields. While these modes are coupled to a thermal reservoir, their finite response time allows us to transiently overcome the deleterious effects of the environmental coupling and generate strong two-mode correlations. Second, in contrast to a conventional interferometer, a nondegenerate parametric amplifier takes the place of the input beamsplitter. As proposed in \cite{yurke1986}, such a configuration, also referred to as a $SU(1,1)$ interferometer, exhibits interferometric sensitivity surpassing the SQL due to the two-mode correlations created by the parametric interaction \cite{gross2010, hudelist2014}. Importantly, in contrast to interferometry with squeezed or entangled input states, the Heisenberg scaling of sensitivity in the $SU(1,1)$ interferometer has been shown to be robust to particle loss and inefficient detection \cite{marino2012}. 

The nonlinear phonon interferometer is described by the interaction Hamiltonian (see SI)
\begin{eqnarray}
 H_{int}(t) &=& i \hbar \frac{\tilde{g}_SX_S(t)}{2} (a_s^\dagger a_i^\dagger - a_s a_i ) \nonumber \\
  &+& i \hbar \frac{\tilde{g}_DX_D(t)}{2} (a_s^\dagger a_i - a_s a_i^\dagger)
 \label{eqn:HInt}
\end{eqnarray}
where $a_s,a_i$ are the annihilation operators of the two interferometric modes, hereafter referred to as the `signal' and `idler' modes, with resonance frequencies $\omega_s$ and $\omega_i$ respectively,  
$\tilde{g}_S,\tilde{g}_D$ are coupling strengths between the two modes at the sum and difference frequencies, and $X_S(t),X_D(t)$ are the amplitudes of the supporting substrate modes (`pump') at the sum and difference frequencies. 
The first term represents the nondegenerate parametric oscillator that causes the correlated production of down-converted phonons in the signal and idler modes. The second term signifies the beamsplitter interaction that results in the coherent exchange of phonons between the signal and idler modes.

\begin{figure}[t]
\begin{center}
\includegraphics[width=2.7in]{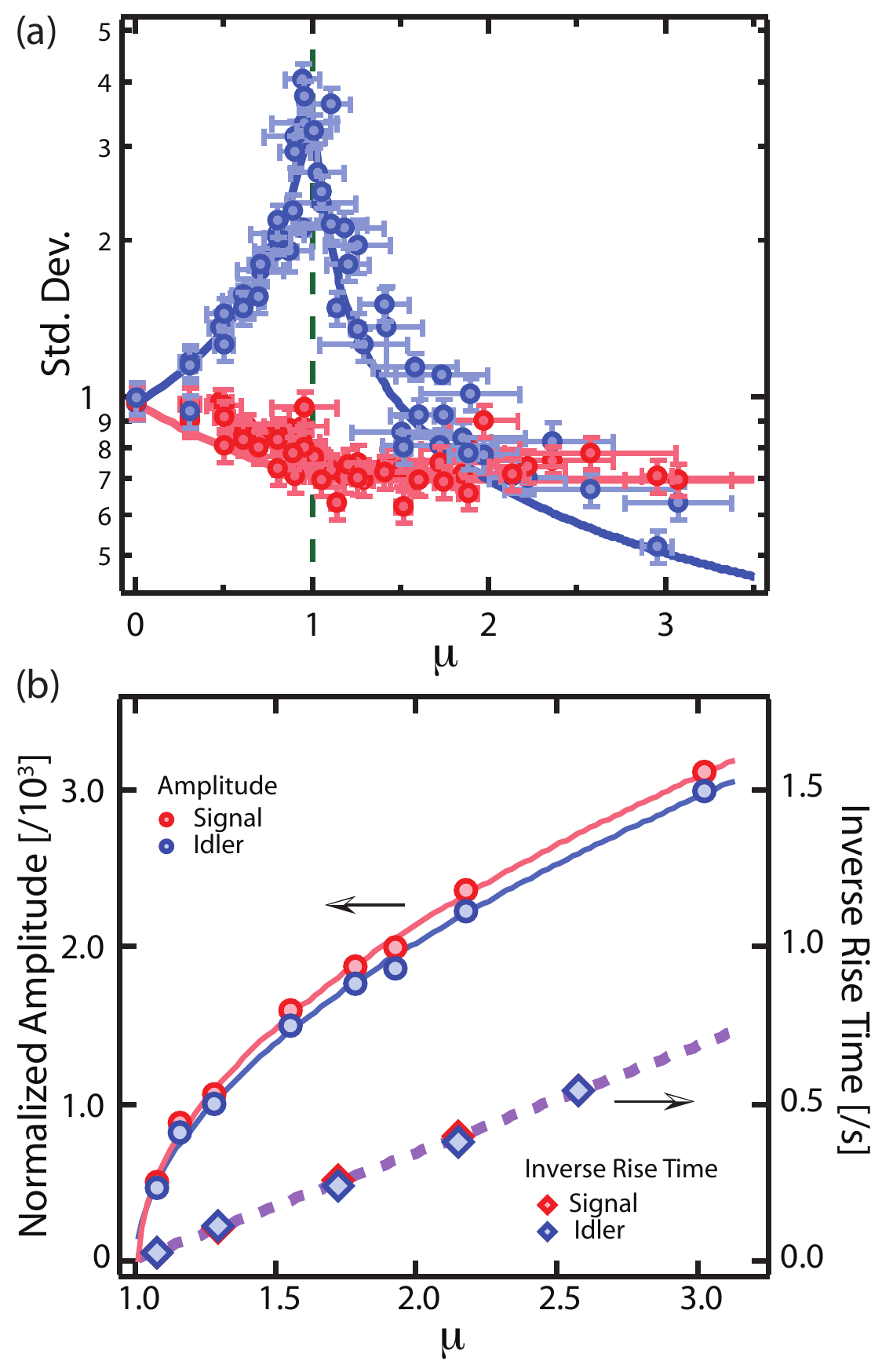}
\caption{\textbf{The parametric amplifier phase diagram.} (a) Two-mode squeezing below and above the instability threshold ($\mu=1$): normalized standard deviations of squeezed (red) and amplified (blue) cross-quadratures. The solid lines are no-free-parameter predictions of our model with independently measured damping rates and eigenfrequencies, taking into account finite measurement time and differential substrate temperature effects (see SI). (b) Steady state amplitudes of the signal and idler modes show a power-law growth of $0.53\pm0.03$ consistent with the prediction of $0.5$. The exponential growth rate of the signal idler motions increases linearly with parametric amplifier actuation $\mu$. (All signal and idler motions are normalized to their respective thermal motions.)}
\label{fig:PA_phase}
\end{center}
\end{figure}

The parametric amplifier and beamsplitter interactions in our system are independently ascertained. For the experiments described below, the resonance frequencies and damping rates of the signal and idler modes are $\omega_s/2\pi = 1.233$ MHz, $\omega_i/2 \pi = 1.466$ MHz and $\gamma_s/2 \pi = 0.083(2)$ Hz, $\gamma_i/2\pi = 0.108(3)$ Hz. As is well known in quantum optics, the parametric amplifier shows an instability when driven past a critical pump amplitude, $X_{S,th}$, where the system is characterized by a divergent mechanical susceptibility and critical dynamics. This instability can be regarded in terms of a nonequilibrium continuous phase transition \cite{gerry1988, dechoum2004}. When the substrate is driven beyond this threshold, the signal and idler modes self-oscillate, achieving a steady state when their decay rate matches the rate of downconversion from the sum frequency parametric drive. 
The strength of the parametric drive can thus be parametrized by $\mu \equiv X_S/X_{S,th}$ with $\mu=1$ representing the critical point for the onset of self-oscillation. Below threshold, the motional amplitudes of the signal and idler modes are zero while their fluctuations are correlated, and the cross-quadratures $\frac{x_s\pm x_i}{\sqrt{2}}, \frac{y_s\pm y_i}{\sqrt{2}}$ are squeezed. Above threshold, the modes get correlated and their amplitude sum, difference $\frac{r_s\pm r_i}{\sqrt{2}}$ are squeezed, where $r_{s,i}=\sqrt{x_{s,i}^2+y_{s,i}^2}$ \cite{art:Chakram2015}. The cross-quadrature and amplitude sum, difference squeezing phase diagram is shown in Fig.~\ref{fig:PA_phase}(a), and shows excellent agreement with a no-free-parameter calculation based on the model and independently measured damping and frequency parameters. The growth of the steady state amplitude of the modes above threshold shown in Fig.~\ref{fig:PA_phase}(b) is measured to have a power-law growth with exponent $0.53\pm0.03$, in close agreement with the theoretical prediction ($1/2$). Also, the exponential growth rate of the modes' amplitudes increases linearly in the drive strength $\mu$, as predicted by our parametric amplifier model. 

\begin{figure*}[t]
\centering
\includegraphics[width=0.850\textwidth]{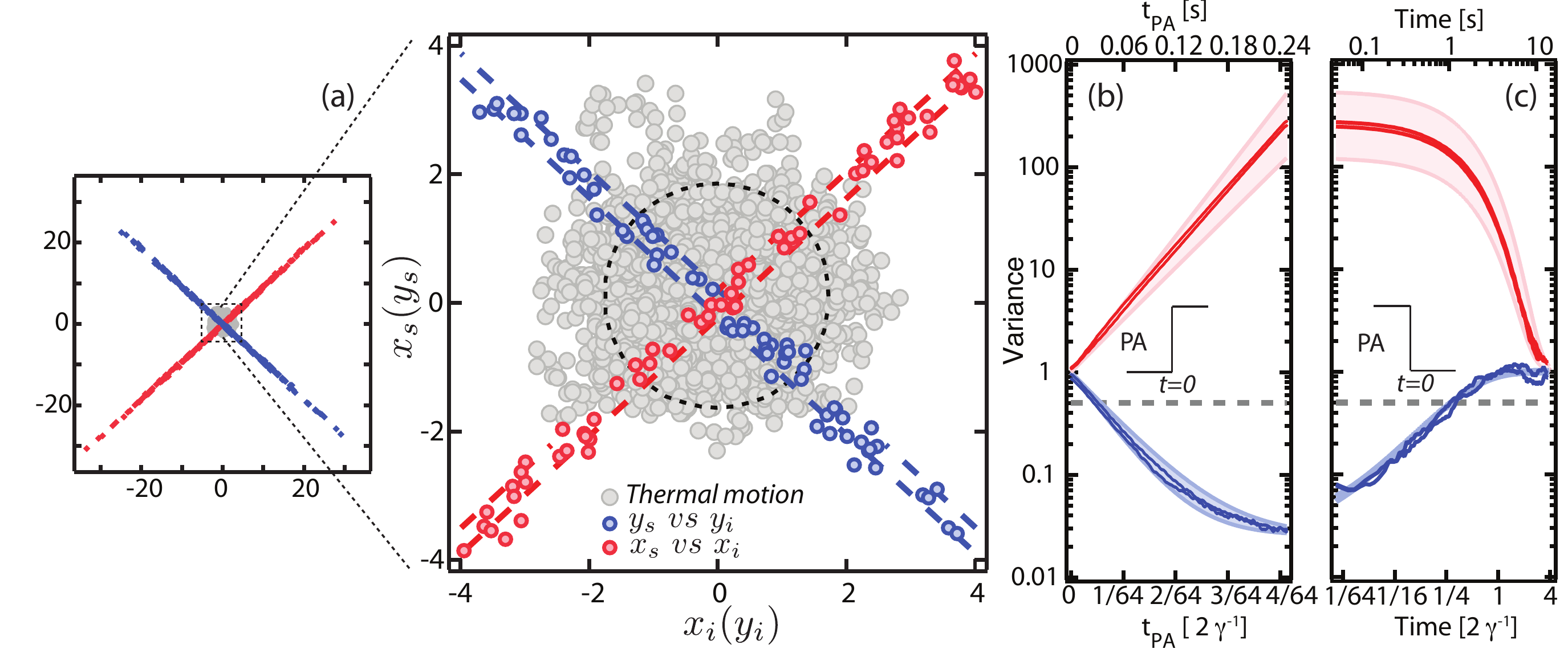}
\caption{\textbf{Enhanced transient squeezing -- phase space distribution and squeezing dynamics.} By transient application of the parametric interaction, two-mode squeezing beyond the steady-state bound of 3 dB can be achieved (see main text).  (a) Phase space distribution of $15.4\pm0.3$dB squeezed states -- $(x_s,x_i)$ (red) and $(y_s,y_i)$ (blue). The thermal state (grey) is shown for reference. (b,c) Dynamics of growth and decay of the two-mode squeezed state in units of the mechanical damping time $2 \gamma^{-1}$. The parametric amplifier is driven transiently with strength $\mu = 38(5)$, and the quadrature variance of 236 iterations is plotted {\em vs} time. For comparison, the steady state bound of 3 dB is indicated in grey. The shaded regions represent no-free-parameter bounds due to variations in the parametric drive $\mu$ across the iterations.  (All signal and idler displacements have been normalized to their respective thermal amplitudes measured at $t=0$.)}
\label{fig:TranSq}
\end{figure*}

The mechanical beamsplitter interaction \cite{art:faust2013, okamoto2013} realizes a coherent transfer of quanta between the signal and idler modes. By making weak optical measurements of the signal and idler modes, this interaction can also be realized in {\em post hoc} analysis. In this work, we perform weak, independent measurements on the signal and idler output modes with minimal backaction, and coherently combine the measured quadratures to effect the coherent beamsplitter. Back-action evading measurements \cite{wollman2015} allow for such a realization of the beamsplitter to be extended into the quantum regime. 

As shown in Fig.~\ref{fig:figschem}(b), the nonlinear interferometer is realized in the time domain. The signal and idler modes are initialized in a coherent product state $|\alpha_s, \alpha_i \rangle$ by independently actuating the two modes at times $t < 0$. At $t=0$, the parametric amplifier is actuated for a duration $t_{PA} \ll \gamma_{s,i}^{-1}$ by parametrically driving the substrate at the sum frequency $\omega_{PA} = \omega_s + \omega_i$. Subsequently, the signal mode interacts with the parameter of interest for a variable duration. Lastly, the beamsplitter interaction is pulsed for a duration $t_{BS} \ll \gamma_{s,i}^{-1}$ by parametrically actuating the substrate at the difference frequency $\omega_{BS} = \omega_s - \omega_i$, and the two output modes are independently measured. 

The dynamical behavior of the phonon interferometer is governed by the equations of motion for the interfering modes given by 
$\dot{a}_{s,i} = \frac{\mu}{2} \sqrt{\gamma_s \gamma_i} a^\dagger_{i,s} - \frac{\gamma_{s,i}}{2} a_{s,i} + \sqrt{\gamma_{s,i}} a_{s,i}^{in}$, where the input fields $a_{s,i}^{in}$ are the coherent input states $|\alpha_s, \alpha_i \rangle$ with unit variance normalized to the mode's thermal motion, i.e. $\Delta X^2_{s,in} = \Delta Y^2_{s,in} = 1$, and we have assumed a Markovian reservoir. While the simultaneous solution to these equations is straightforward for the experimentally relevant case of mismatched damping rates ($\gamma_i \neq \gamma_s$) \cite{art:Chakram2015}, we state the results for identical damping rates and note that the conclusions remain essentially unaltered (see SI). For $\mu \gg 1$, the output variances are minimized for a beamsplitter mixing angle $\phi = \tilde{g}_D X_D t_{BS}/2 = -\pi/4$ and are respectively given by $\langle \Delta X^2_{out} \rangle = \frac{1}{1 + \mu} + \frac{\mu}{1 + \mu} e^{-\gamma (1 + \mu) t_{PA}}$ and $\langle \Delta Y^2_{out} \rangle = \frac{1}{1 - \mu} + \frac{\mu}{\mu - 1} e^{\gamma (\mu - 1) t_{PA}}$ where $X_{out} \equiv \frac{x_s - x_i}{\sqrt{2}}$, $Y_{out} \equiv \frac{x_s + x_i}{\sqrt{2}}$ are the output quadratures. The squeezed $X$-quadrature reduces exponentially to a variance $1/(1+\mu)$ with a time constant $[\gamma (1 + \mu)]^{-1} \ll \gamma^{-1}_{s,i}$ allowing for a significant noise squeezing well beyond the 3 dB bound. By seeding the modes with only thermal motion, i.e. $\alpha_s,\alpha_i=0$, we demonstrate these dynamics and a resultant noise reduction by $15.4\pm0.3$ dB, (Fig.~\ref{fig:TranSq}). The slow decay to the thermal state, shown in Fig.~\ref{fig:TranSq}c, enables an unbounded improvement of SNR for the detection of impulsive forces or the detection of forces at enhanced bandwidth. 

\begin{figure}
\begin{center}
\includegraphics[width=3.25in]{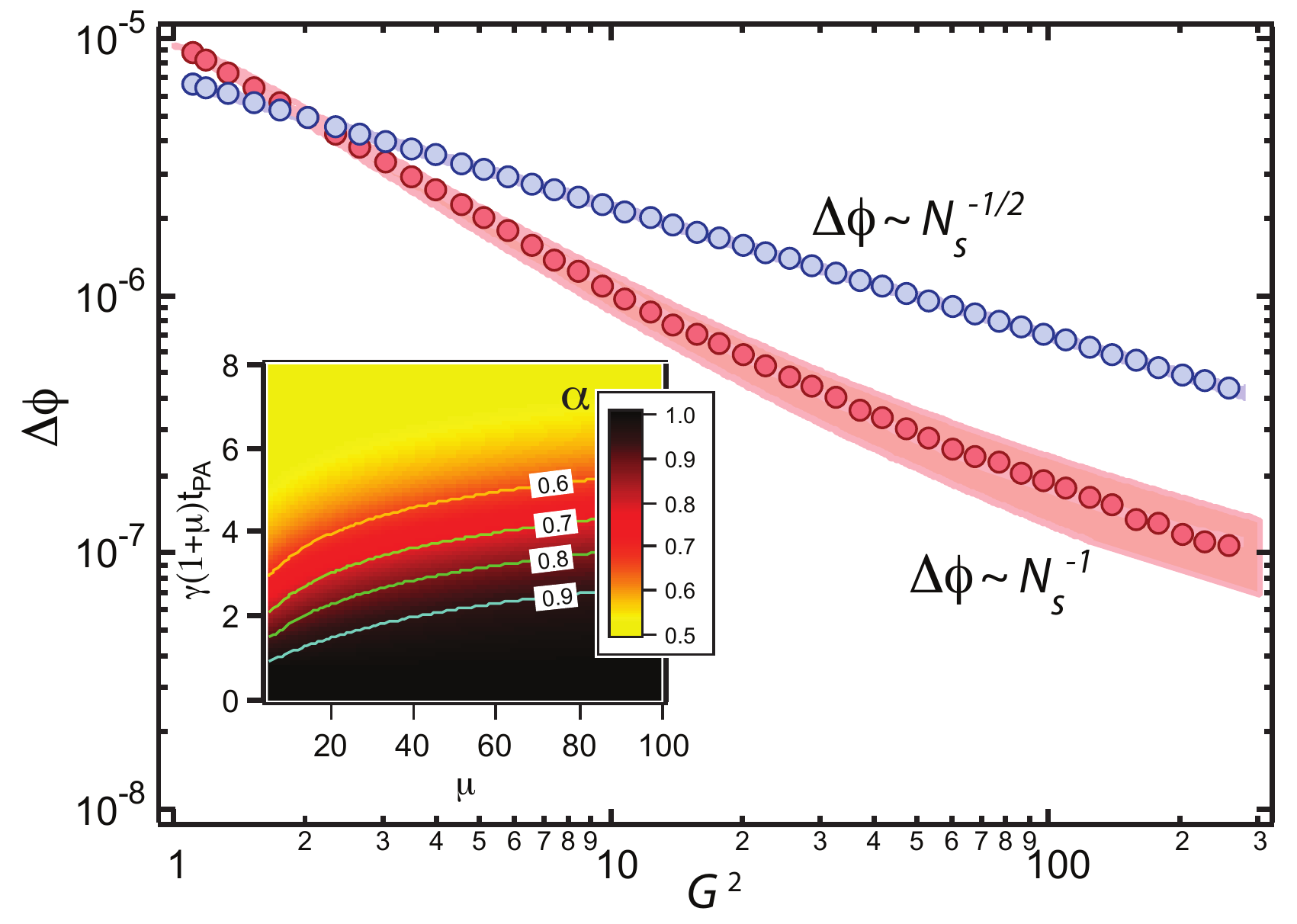}
\caption{The Heisenberg scaling of phase sensing in the nonlinear phonon interferometer is shown {\em vs} the phonon number gain $G^2$. The shot noise limit for conventional interferometry, i.e. in the absence of two-mode correlations, is shown for comparison (blue). The data correspond to the experimental parameters of Fig.~\ref{fig:TranSq}. The shaded regions represent no-free-parameter bounds due to variations in the parametric drive $\mu$ across the iterations.
Inset: The phonon interferometer's estimated scaling exponent $\alpha$ for the phase sensitivity, $\delta \phi \sim 1/N^\alpha$, is shown as a function of the parametric drive $\mu$ and the parametric pulse duration $t_{PA}$, indicating the transition from SQL scaling ($\alpha = 1/2$) to Heisenberg scaling ($\alpha = 1$) as $t_{PA}$ is reduced (see text). }
\label{fig:Sens}
\end{center}
\end{figure}

Further, the output $X$-quadrature for a coherent input state $|\alpha_s, 0 \rangle$ can be evaluated from the above equations of motion to be $\langle X_{out} \rangle = \frac{\operatorname{Re}[\alpha_s]}{2} [ (\cos \phi - \sin \phi) e^{-\gamma (\mu + 1) t_{PA}/2} + (\cos \phi + \sin \phi) e^{\gamma (\mu - 1) t_{PA}/2}]$. 
For a parametric pulse duration $t_{PA}$ such that $t_{PA} \ll \gamma^{-1}$, the remnant noise in this quadrature is $\langle \Delta X^2_{out} \rangle^{1/2} \propto e^{-\gamma (\mu+1) t_{PA}}$ (see SI) and the minimum detectable phase is 
\begin{equation}
\Delta \phi = \frac{\langle\Delta X_{s,out}^2\rangle^{1/2}}{d\left<X_{s,out}\right>/d(\phi)}\propto\frac{\operatorname{Re}[\alpha_s]}{N_s}
\end{equation}
where $N_s \approx (1/4) (\operatorname{Re}[\alpha_s] e^{\gamma \mu t_{PA}/2})^2 = (1/4) (\operatorname{Re}[\alpha_s] \times G)^2$ is the mean phonon number in the signal mode, showing the Heisenberg scaling of phase sensitivity with measurement resource (Fig.~\ref{fig:Sens}). As indicated by the dynamical equations for the interferometric modes, the degree of two-mode squeezing saturates as the parametric pulse duration $t_{PA}$ approaches the mechanical damping time $\gamma^{-1}$ and the nonlinear interferometer reverts back to SQL scaling (Fig.~\ref{fig:Sens},inset). 

In conclusion, we demonstrate a $SU(1,1)$ phonon interferometer capable of Heisenberg-limited phase sensitivity using parametrically coupled mechanical modes in a monolithic SiN membrane resonator. Owing to the large $f \times Q$ product of the interferometric modes, we demonstrate a substantial degree of transient two-mode squeezing, achieving a noise reduction of $15.4(3)$ dB, well beyond the conventional $3$ dB bound, and show that this transient state is long lived, surviving on the order of $10^6$ mechanical periods. Our work extends the optomechanical toolbox for the quantum manipulation of macroscopic mechanical motion, and enables new techniques for optomechanical sensing and the manipulation of mechanical fluctuations. Extending these techniques to the quantum regime should allow for studies of macroscopic decoherence in highly correlated phononic states. Even in the classical regime, we note that the emergence and decay of two-mode correlations and the ensuing thermalization dynamics are intimately tied to the nature of the reservoir that couples to the interferometric modes. As such, ultraprecise phonon interferometers such as demonstrated in this work enable the study of non-equilibrium optomechanical dynamics, the interferometric detection of  non-Markovian dynamics \cite{groblacher2015} arising from non-Ohmic reservoirs, and the harnessing of such reservoirs for the creation and stablization of macroscopic non-classical states \cite{ramos2013}. 

\textbf{Acknowledgments} 
This work was supported by the DARPA QuASAR program through a grant from the ARO, the ARO MURI on non-equilibrium Many-body Dynamics (63834-PH-MUR) and an NSF INSPIRE award. M. V. acknowledges support from the Alfred P. Sloan Foundation.

\textbf{Contributions} H. F. H. C., Y. S. P., L. C. and S. C.  performed the experimental work and data acquisition. H. F. H. C. and Y. S. P. carried out the data analysis and modelling. 
M.V. supervised all stages of the work. All authors contributed to the preparation of the manuscript.

\textbf{Correspondence} Correspondence should be addressed to M.V.~(mukundv@cornell.edu).
\newpage
\section{SUPPLEMENTARY INFORMATION}

\subsection{Optical detection and stabilization of mechanical modes}

The mechanical resonators studied in this work are the eigenmodes of a square silicon nitride membrane resonator fabricated by NORCADA Inc. The membranes have lateral dimensions of 5 mm and a thickness of 100 nm, with typical mechanical quality factors in the range of $10^7$ \cite{chakram2014}. The displacement of the membrane modes are detected using a Michelson optical interferometer with a position sensitivity of 0.1 pm/Hz$^{1/2}$ for typical powers of 200 $\mu$W incident on the membrane. An external cavity diode laser operating at a wavelength of 795 nm provides the light for this interferometer. 

Due to the differential thermal expansion of the membrane and the supporting substrate, the mechanical modes of the resonators are susceptible to large frequency drifts due to temperature fluctuations. Thus, the precise measurement of thermomechanical motion and non-thermal two-mode correlations requires active sub-linewidth stabilization of the mechanical eigenfrequencies. This is accomplished by photothermal control of the silicon substrate. As described in previous work \cite{patil2015}, we implement active stabilization by continuously monitoring the mechanical eigenfrequency of a high-$Q$ membrane mode at 2.736 MHz - far from the modes of interest in this work. Phase sensitive detection of this mode generates an error signal with an on-resonant phase slope of 5.91 radians/Hz. Active photothermal stabilization of the substrate is accomplished with typical powers of 600 $\mu$W, generated by a diode laser at 830 nm. Under this active stabilization, the rms frequency fluctuations of this `thermometer mode' are measured to be below 2 mHz, equivalent to temperature fluctuations of the substrate of less than 2 $\mu$K. For the modes of the interferometer, this translates to frequency fluctuations less than $0.002 \times \gamma$. 

At low frequencies ($< 3$ Hz), the optical interferometer used for the detection of mechanical displacement is susceptible to residual amplitude modulation (RAM) due to gradual temperature fluctuations and temperature-dependent birefringence of the various optical elements. In our experiments, this low frequency amplitude noise (measured to be around 10 ppm with respect to the carrier peak) convolves with the mechanical displacement signal leading to a 0.75\% contamination of the detected membrane displacement and the two-mode correlations. Due to this low-frequency RAM, the signal-to-noise ratio for the mechanical thermal motion degrades to 1 as the mechanical amplitude approaches 150 times the room-temperature thermal amplitude. This restricts the dynamic range of membrane amplitudes over which we can study the mechanical interferometer. In this work, we overcome this limitation by artificially increasing the mechanical thermal noise (and effective temperature of the mechanical modes), by driving each mode of the phonon interferometer with gaussian noise centered at each mechanical eigenfrequency over a bandwidth of 10 Hz, much larger than the respective mechanical linewidths. 
Techniques of active RAM control \cite{zhang2014b} can be used to extend our measurements to the quantum regime of mechanical motion. 

\subsection{Parametric amplifier and beam splitter dynamics}
The two-mode nonlinearities are described by an interaction Hamiltonian \cite{patil2015}
\begin{equation}
H_{int}(t) = -g_{S,D} X_{S,D}(t) x_s x_i
\label{eqn:HInt0}
\end{equation}
where $g_{S,D}$ parametrize the strength of the interactions, $X_{S,D}$ are the amplitudes of the Silicon substrate excitations at the sum and difference frequencies, and $x_{s,i}$ are the amplitudes of the individual membrane resonator modes. Defining $a_{s,i}$ as the annihilation operators of the signal and idler modes, their motional amplitudes are given by $x_{s,i}= x_{(s,i),0}(a_{s,i}+a_{s,i}^{\dagger})$ where $x_{(s,i),0}= (\hbar/2m_{s,i}\omega_{s,i})^{1/2}$ are their respective zero point motions. On normalizing the  coupling constants as $\tilde g_{S,D} = g_{S,D} x_{s,0}x_{i,0}/2\hbar$, the interaction Hamiltonian under the rotating wave approximation simplifies to
\begin{eqnarray}
 H_{int}(t) &=& i\hbar \frac{\tilde g_SX_S(t)}{2} (a_s^\dagger a_i^\dagger - a_s a_i ) \nonumber \\
 	&+& i \hbar\frac{\tilde g_DX_D(t)}{2} (a_s^\dagger a_i - a_s a_i^\dagger)
 \label{eqn:HInt}
\end{eqnarray}
as in the main text. When only the parametric amplifier coupling is pulsed on for time $t_{PA}$, the evolution of the fields is governed by the squeezing Hamiltonian given by the first term. In the Heisenberg picture, $a_{s,i}$ evolve as $a_{s,i}(t)=Ga_{s,i}(0)+ga_{i,s}^{\dagger}(0)$, where $G=\cosh(\tilde g_SX_S  t_{PA}/2)$ and $g=\sinh(\tilde g_SX_S  t_{PA}/2)$ parametrize the pulsed parametric amplifier gains. The dynamics when the beam splitter coupling is pulsed on for time $t_{BS}$ is governed by the second term, and $a_{s,i}$ evolve at final time $t_f$ into $a_{s,i}(t_f)=\cos \phi \, a_{s,i}(t_{PA})\pm \sin \phi \, a_{i,s}(t_{PA})$, where the mixing angle $\phi = \tilde g_DX_D t_{BS}/2$ characterizes the coherent mixing between the signal and idler modes due to the beam splitter interaction. 

The output variance of the $X$-quadrature of the signal mode $\Delta X_{s,out}^2$ (or of the $Y$-quadrature of the idler mode $\Delta Y_{i,out}^2$) depends on the mixing angle $\phi$, where the quadratures are defined such that $a_{s,i}=X_{s,i}+iY_{s,i}$ . For the coherent states $|\alpha_s,\alpha_i\rangle$ which form the input to the interferometer, the variances are $(\Delta X_{(s,i)}^{in})^2=(\Delta Y_{(s,i)}^{in})^2=1$, where the motion of each mode is normalized to its thermal or quantum zero-point motion. For $G,g\gg1$, the output variances are minimized for $\phi = -\pi/4$ and equals $\Delta X_{s,out}^2(t_f)=\Delta Y_{i,out}^2(t_f)=1/(G+g)^2=\exp(-\tilde g_SX_S t_{PA})$. This improves the SNR of measuring the phase $\phi$ by an exponential factor $\exp(-\tilde g_SX_S t_{PA}/2)$ as compared to that achieved using a conventional interferometer, for which $\Delta X_{s,out}^2(t_f)=\Delta Y_{i,out}^2(t_f)=1$. This improvement is, in principle, unbounded - the parametric amplifier gains $G\approx g=\sinh(\tilde g_SX_S t_{PA}/2)$ diverge on increasing the argument. 

Unlike in the optical domain, phononic fields are necessarily confined in a cavity -- the mechanical resonator -- and do not propagate freely without losses. The lossy system can be formally modelled as a linear coupling of the mechanical modes to an environmental bath of harmonic oscillators. For a Markovian bath, the dynamics can be evaluated using the input-output formalism, and for the case of matched frequencies and loss rates is governed by $\dot a_{s,i}=\frac{\mu}{2}\gamma a_{i,s}^\dagger -\frac{\gamma}{2}a_{s,i}+\sqrt{\gamma}a_{s,i}^{in}$, where $\mu = X_S/X_{S,th}$ parametrizes the parametric pump drive strength at the sum frequency. 
Defining the cross-quadrature modes $d_\pm=(a_s\pm a_i)/\sqrt{2}$, the equations of motions are rewritten as $\dot X_{d_+}=\frac{\gamma}{2}(\mu-1) X_{d_+}+\sqrt{\gamma}X_{d_+}^{in}$ and $\dot Y_{d_+}=-\frac{\gamma}{2}(\mu+1) Y_{d_+}+\sqrt{\gamma}Y_{d_+}^{in}$, and similarly for $X_{d_-},Y_{d_-}$. Assuming an initial thermal or quantum vacuum seed, the dynamics of their variances is evaluated to be $\langle \Delta Y_{d_+}^2(t)\rangle =  \langle \Delta X_{d_-}^2(t)\rangle =  \frac{1}{1+\mu}+\frac{\mu}{1+\mu}e^{-\gamma(\mu+1)t}$, which are squeezed, and $\langle \Delta X_{d_+}^2(t)\rangle =  \langle \Delta Y_{d_-}^2(t)\rangle =  \frac{1}{1-\mu}+\frac{\mu}{\mu-1}e^{\gamma(\mu-1)t}$, which are amplified.

For a beam splitter interaction with mixing angle $\phi$, the signal output quadratures are expressed in terms of the cross-quadratures as $X_{out}=\sin(\phi+\pi/4)X_{d_+}+\cos(\phi+\pi/4)X_{d_-}$ and $Y_{out}=\sin(\phi+\pi/4)Y_{d_+}+\cos(\phi+\pi/4)Y_{d_-}$.
The variance of $X_{out} $ is minimized for a mixing angle $\phi=-\pi/4$, i.e. when the quadratures are $X_{out}=X_{d_-}$ and $Y_{out}=Y_{d_-}$, and the dynamics of their variances is $\langle \Delta X_{out}^2\rangle=\frac{1}{1+\mu}+\frac{\mu}{1+\mu}e^{-\gamma(1+\mu)t_{PA}}$ and $\langle \Delta Y_{out}^2\rangle=\frac{1}{1-\mu}+\frac{\mu}{\mu-1}e^{\gamma(\mu-1)t_{PA}}$, where the inputs $a_{s,i}^{in}$ are assumed to be coherent states $|\alpha_{s,i}\rangle$. While these dynamics are exact below threshold ($\mu<1$), their validity well above threshold, as in this work, is restricted to small times $t_{PA}<\tau_{s,i}$ (see ``Pump Depletion" below). Importantly, because the squeezed quadrature variances reduce exponentially to $1/(1+\mu)$ with a time constant $[\gamma(1+\mu)]^{-1}\ll\tau_{s,i}$, there is ample time for the squeezed variances to break the steady state squeezing bound of $3$ dB \cite{art:Chakram2015,patil2015}.

\subsection{Pump Depletion}

The assumption that $\mu(t)=\mu$ is a constant over the course of evolution of the parametric amplifier pulse is invalid for $\mu_0\equiv\mu(t=0)>1$. As the signal and idler amplitudes grow exponentially, the absolute downconverted phonon loss rate increases before balancing out with the absolute downconversion rate, thereby decreasing $\mu(t)=\frac{X_S(t)}{X_{S,th}}$. This effect, which arises from the interference of the signal, idler and pump modes, is referred to as pump depletion.

To quantify the effect of pump depletion, we first consider the equations of motion of the signal, idler and sum-frequency pump modes. Within the rotating wave approximation, the Hamiltonian (\ref{eqn:HInt0}) gives 
\bea
\ddot{x}_{s,i} &+& \gamma_{s,i} \dot{x}_{s,i} + \omega_{s,i}^2 x_{s,i} = \frac{1}{m_{s,i}} (F_{s,i}(t) + \frac{g_S}{2} X_S x^*_{i,s})\label{fasts}\nonumber\\
\ddot{X}_S &+& \gamma_S \dot{X}_S + \omega_S^2 X_S = \frac{1}{m_S} (F_S(t) + \frac{g_S}{2} x_s x_i)\label{fastS}
\eea
where $x_{s,i}, X_S$ denote the complex displacement of each mode and $m_{s,i,S},\ \omega_{s,i,S},\ \gamma_{s,i,S}$ and $F_{s,i,S}$ denote respectively the masses, frequencies, damping rates and forces on each of the modes. These coupled equations of motion can be solved using two timescale perturbation theory \cite{lifshitz2008} and simplify to the first order coupled equations,
\bea
2 \dot{A}_{s}  &=& \gamma_{s} \left[ - A_{s}  + i \frac{g_S}{2}\chi_{s}A^*_{i} A_S +  i\chi_{s}\tilde{F}_{s}(t)\right]\label{slows} \nonumber\\
2 \dot{A}_{i}  &=& \gamma_{i} \left[ - A_{i}  + i \frac{g_S}{2}\chi_{i}A^*_{s} A_S +  i\chi_{i}\tilde{F}_{i}(t) \right]\label{slowi} \nonumber\\
2 \dot{A}_{S}  &=& \gamma_{S} \left[ - A_{S} + i \frac{g_S}{2}\chi_{S}A_{s} A_{i} +  i\chi_{S}\tilde{F_{S}}(t)\right]\label{slowS}
\eea
where $x_k = A_k e^{-i \omega_k t},\,\, k\in[i,s,S]$; $\tilde{F}_{k}$ are the slowly varying (complex) amplitudes of the external forces on the individual modes, and $\chi_{k} = (m_k \omega_k \gamma_k)^{-1}$ are their on-resonant susceptibilities. We ignore terms such as $\ddot{A}_k, \gamma_i \dot{A}_k$ in the slow time approximation.
These are further simplified to
\begin{eqnarray}
2\dot{ \tilde{ A_s}}&=\gamma_s[-\tilde A_s +i \tilde A_i^*\tilde A_S +i  \mathcal{F}_s(t)]\nonumber\\
2\dot {\tilde {A_i}}&=\gamma_i[-\tilde A_i +i \tilde A_s^*\tilde A_S +i  \mathcal{F}_i(t)]\nonumber\\
2\dot {\tilde {A_S}}&=\gamma_S[-\tilde A_S +i \tilde A_s \tilde A_i +i \mathcal{F}_S(t)]
\end{eqnarray}
where the motion of the pump $\tilde A_S$ is normalized with respect to $A_{S,th}$ -- the critical value which defines the instability threshold $\mu=1$, $\tilde A_{s,i}$ are normalized with respect to their characteristic steady-state motion above threshold $\frac{2}{g_S\sqrt{\chi_S\chi_{i,s}}}$, and $\mathcal{F}_k$ are normalized forces on the respective modes, with $\mathcal{F}_S(t)=\mu(t)$. Lastly, we assume the empirical fact that the pump mode, which is a Silicon substrate mode, has a much larger damping rate than the signal and idler resonator modes,  i.e. $\gamma_S\sim (10^3-10^4) \gamma_s,\gamma_i$, thereby ensuring that the pump motion adiabatically follows the signal and idler motions, i.e. $\tilde A_S=i \tilde A_s \tilde A_i +i \mathcal{F}_S(t)$. 

$\tilde A_s$ and $\tilde A_i$ increase exponentially as $\exp(\gamma(\mu_0-1)t/2)$ after the parametric amplifier pulse is switched on, causing $\mu(t)$ to deplete and settle to a steady state of $\mu(t)\xrightarrow{t\rightarrow\infty}1$ \cite{art:Chakram2015}. For $\mu_0>1$, the transient squeezing expression derived assuming constant  $\mu(t)=\mu_0$ is valid only for small times $t_{PA}<\tau_{s,i}$ (see also \cite{art:wolinsky1985,art:carmichael1986}). Considering a case where the seed motions are $\tilde A_{s0}$ and $\tilde A_{i0}$, a pump depletion by a factor $\eta$ occurs when
\begin{equation}
|\tilde A_s \tilde A_i|=\eta|F_S|=\eta\mu_0
\end{equation}
Dropping the absolute modulus sign for clarity, this occurs when
\begin{eqnarray}
\tilde A_{s0} \tilde A_{i0}e^{\gamma(\mu_0-1)t_{PA}}=\eta\mu_0 \nonumber
\end{eqnarray}
For time $t_{PA}\sim\frac{\ln(10\mu_0)}{\gamma(1+\mu_0)}$, the squeezed variances reduce to within $10$\% of $\frac{1}{1+\mu_0}$, and for this time $t_{PA}$,
\begin{equation}
\eta\sim (10\mu_0)^{\frac{\mu_0-1}{\mu_0+1}}\frac{\tilde A_{s0}\tilde  A_{i0}}{\mu_0} \xrightarrow{\mu_0\gg1}10 \tilde A_{s0} \tilde A_{i0}
\end{equation}
For the motional amplitudes in the present study (data in Figs(3,4) of the main text), $\tilde A_{s0}\approx \tilde A_{i0}\approx0.03$, and $\eta\approx0.01$, i.e. there is a mere $1\%$ pump depletion even at almost saturated squeezing.

Clearly, the degradation of squeezing due to pump depletion sets in sooner for (i) larger seeds $\tilde A_{s0},\tilde A_{i0}$, (ii) larger damping of the signal, idler and pump modes, and (iii) larger nonlinear signal-idler couplings $\tilde g_S$ (see also \cite{art:carmichael1986}). We note that the deleterious effects of pump depletion can be avoided by compensating the pump depletion with feedback and other control systems.

\subsection{Heisenberg-scaling of phase sensitivity}
The interferometer output $X$-quadrature for a coherent input state $|\alpha_s,0\rangle$ can be calculated from the above equations of motion to be 
\begin{eqnarray}
\left<X_{s,out}\right>&=&\frac{Re[\alpha_s]}{\sqrt{2}}\bigg(\frac{\cos \phi-\sin \phi}{\sqrt{2}}e^{-\gamma(\mu+1)t_{PA}/2} \nonumber \\
 &+& \frac{\cos \phi+ \sin \phi}{\sqrt{2}}e^{\gamma(\mu-1)t_{PA}/2}\bigg) \\
\frac{d\left<X_{s,out}\right>}{d(\phi)} &=& \frac{Re[\alpha_s]}{\sqrt{2}}\bigg(\frac{-\sin \phi-\cos \phi}{\sqrt{2}}e^{-\gamma(\mu+1)t_{PA}/2} \nonumber \\
	&+& \frac{-\sin \phi+\cos \phi}{\sqrt{2}}e^{\gamma(\mu-1)t_{PA}/2}\bigg)
\end{eqnarray}

For large phonon gain of the parametric amplifier, the effects of pump depletion cause a decrease in noise squeezing and a reversion of the interferometer to SQL scaling ($\delta \phi \sim 1/\sqrt{N_s})$. As such, the choice of the parametric amplifier duration $t_{PA}$ is not independent of the parametric drive $\mu$. To take this mutual dependence into account, we parametrize the amplifier duration as $t_{PA}=\frac{\log[k\mu]}{\gamma(1+\mu)}$, where $k>1/\mu$ is a dimensionless parameter. With this parametrization, the remnant noise and the minimum detectable motion $\Delta X_D$ is thus given by
\begin{eqnarray}
\langle\Delta X_{s,out}^2\rangle^{1/2}=\sqrt{k+1}e^{-\gamma(\mu+1)t_{PA}/2}\nonumber\\
\Delta \phi =\frac{\langle\Delta X_{s,out}^2\rangle^{1/2}}{d\left<X_{s,out}\right>/d\phi}\propto\frac{\operatorname{Re}[\alpha_s]}{N_s}
\end{eqnarray}
where $N_s \approx (\operatorname{Re}[\alpha_s]e^{\gamma\mu t_{PA}/2})^2/4 = G^2 \times (\operatorname{Re}[\alpha_s]/2)^2$ is the mean phonon number in the signal mode, which is the measurement resource. Here, we assume that  $\operatorname{Re}[\alpha_s]\gg1$ and $e^{\gamma\mu t_{PA}}\gg1$, and define $G^2 = e^{\gamma \mu t_{PA}}$ to be the phonon number gain. 

\subsection{Effect of finite substrate temperature}
To reduce the contribution of residual amplitude modulation in the optical interferometer to the signal and idler motion readout, all data are acquired by artificially driving the signal and idler modes to an elevated temperature, corresponding to an effective thermal motion around 40-50 times the room temperature amplitude, by driving the two modes with a gaussian noise source with a bandwidth about 100 times larger than the respective mechanical linewidths. The substrate mode, however, is not artificially driven to a larger temperature. 

The substrate mode fluctuations couple equally to the signal and idler mode, and thus the amplitude-difference ($\frac{x_s-x_i}{\sqrt{2}}$) squeezing is independent of and robust to any substrate fluctuations. However, substrate fluctuations couple to the amplitude-sum quadrature ($\frac{x_s+x_i}{\sqrt{2}}$). Thus with equal substrate, signal, and idler temperatures, the amplitude-sum variance is bounded at $1/2$ in the limit of large parametric drive strengths, $\mu\gg1$. However, if the substrate temperature is much \emph{lower} than the signal and idler temperatures, the substrate fluctuations are negligible in comparison, and the amplitude-sum squeezing beats the conventional 3dB limit -- in fact, the amplitude-sum variance equals $\frac{1}{2(\mu-1)}$. This difference in the signal and idler and substrate temperatures has been accounted for in Fig.~2a of the main text.

\subsection{Effect of finite measurement time}
To accurately model the measurements described in this work, the effects of finite measurement duration need to be considered. This is particularly relevant for the high-$Q$ resonators used in this work. In the vicinity of the critical point $\mu=1$, the mechanical parametric amplifier exhibits a divergent response time that results in extremely long thermalization times ($\sim 10^4 - 10^5$ seconds). Thus, for typical measurement durations in this work ($\sim 100$ seconds), the measured squeezing spectra can deviate appreciably from those computed in steady state. 

In our model, we take the finite measurement duration into account by computing the variances measured over a measurement time $\tau_m$ using the truncated integral of the relevant spectral density as indicated below \cite{art:Chakram2015},
\begin{equation}
\boldsymbol{\sigma}_{\alpha,\beta} = 2 \int_{2 \pi/\tau_m}^{\infty}\mathbf{S}_{\alpha,\beta}(\omega)d\omega
\end{equation}

\subsection{Transient squeezing in the presence of mismatched damping rates}
In the case of mismatched frequencies and damping rates for the signal and idler modes, the main differences are (i) a small deviation of the optimal beam splitter mixing angle away from $g_D X_D t_{BS} = -\pi/4$, (ii) the minimum variance of the squeezed quadrature becomes mismatch dependent, but agress with the expression in the main text 
to within 10\% for asymmetry parameters $|\delta_\omega - \delta_\gamma| < 0.20$ where $\delta_\omega = \frac{\omega_s - \omega_i}{\omega_s + \omega_i}$, $\delta_\gamma = \frac{\gamma_s - \gamma_i}{\gamma_s + \gamma_i}$ \cite{art:Chakram2015}, and (iii) the time constants of transient evolution of the squeezed and amplified quadratures are respectively altered to $|\frac{\gamma_s + \gamma_i}{2} (1 \pm \mu \sqrt{1 - \delta_\gamma^2})|^{-1}$ for $\mu \gg \frac{\delta_\gamma}{\sqrt{1 - \delta_\gamma^2}}$.

\bibliographystyle{unsrt}
\bibliography{Ref.bib,SU11bib}

\end{document}